\documentclass[aps,showpacs,twocolumn,preprintnumbers,multicol]{revtex4}
\usepackage{epsfig}
\usepackage{color}
\usepackage{amssymb,amsmath}

\begin{document}

\setlength{\voffset}{0.cm}
\setlength{\evensidemargin}{0cm}
\setlength{\oddsidemargin}{0cm}
\setlength{\textwidth}{16.25cm}
\setlength{\textheight}{23.5cm}
\setlength{\floatsep}{0pt}
\setlength{\parskip}{1mm}

\newcommand{\red}[1]{\textcolor{red}{#1}}
\newcommand{\Frac}[2]{\frac{\displaystyle #1}{\displaystyle #2}}

\newcommand{\GeV}{~{\rm GeV}}
\newcommand{\MeV}{~{\rm MeV}}
\newcommand{\im}{{\rm Im}}
\newcommand{\re}{{\rm Re}}
\newcommand{\be}{\begin{equation}}
\newcommand{\ee}{\end{equation}}
\newcommand{\bea}{\begin{eqnarray}}
\newcommand{\eea}{\end{eqnarray}}
\newcommand{\beq}{\begin{equation}}
\newcommand{\eeq}{\end{equation}}
\newcommand{\ba}{\begin{array}}
\newcommand{\ea}{\end{array}}
\newcommand{\beqa}{\begin{eqnarray}}
\newcommand{\eeqa}{\end{eqnarray}}
\newcommand{\dis}{\displaystyle}
\newcommand{\cL}{{\cal L}}
\newcommand{\cA}{{\cal A}}
\newcommand{\cO}{{\cal O}}
\newcommand{\da}{^\dagger}
\newcommand{\no}{\nonumber}
\newcommand{\lsim}{\stackrel{<}{_\sim}}
\newcommand{\gsim}{\stackrel{>}{_\sim}}
\newcommand{\ket}[1]{\vert {#1} \rangle}
\newcommand{\bra}[1]{\langle {#1}}
\newcommand{\la}{\langle}
\newcommand{\ra}{\rangle}
\newcommand{\ccdot}{ \! \cdot \! }
\definecolor{rosso}{cmyk}{0,1,1,0.4}
\definecolor{rossos}{cmyk}{0,1,1,0.55}
\definecolor{rossoc}{cmyk}{0,1,1,0.2}
\definecolor{viola}{cmyk}{0,1,0,0.6}
\definecolor{blu}{cmyk}{1,1,0,0.3}
\definecolor{blus}{cmyk}{1,1,0,0.6}
\definecolor{bluc}{cmyk}{1,1,0,0.1}
\definecolor{blul}{cmyk}{0.2,0,0.3,0.6}
\definecolor{bluaz}{cmyk}{1,0.9,0.1,0.3}
\definecolor{bluazz}{cmyk}{1,0,0,0.3}
\definecolor{verde}{cmyk}{0.92,0,0.59,0.25}
\definecolor{verdec}{cmyk}{0.92,0,0.59,0.15}
\definecolor{verdes}{cmyk}{0.92,0,0.59,0.4}
\definecolor{verdel}{cmyk}{0.92,0.2,0.39,0.34}
\definecolor{giallo}{cmyk}{0,0,1,0}
\definecolor{gialloverde}{cmyk}{0.44,0,0.74,0}
\definecolor{arancia}{cmyk}{0.1, 0.56, 0.60, 0.1}

\newpage
\title{Form factor in  $K^+\rightarrow\pi^+\pi^0\gamma$:
interference versus direct emission}

\author{ Luigi~Cappiello and Giancarlo~D'Ambrosio}

\affiliation{
INFN -- Sezione di Napoli\\
{\rm and}\\
Dipartimento di Scienze Fisiche, Universit\`{a} di Napoli
``Federico II''\\
Via Cintia, 80126 Napoli, Italy\\
}
\date{\today}

\begin{abstract}
We analyze the effect of a
 form factor in the magnetic contribution to
 $K^+\rightarrow\pi^+\pi^0\gamma$.
We emphasize how this can show up experimentally: in particular
we try to explore the difference between a possible interference contribution
and a form factor in the magnetic part. The
form factor used for
 $K^+\rightarrow\pi^+\pi^0\gamma$
is analogous to the one for
$K_L\rightarrow\pi^+\pi^-\gamma$, experimentally well established. 
\end{abstract}

\maketitle



\section{Introduction}
Non-leptonic kaon decays
are an important tool to study
weak interactions \cite{GI98, Rafael, BBL96, ENP94,D'Ambrosio:1997ta, BK00,Sozzi:2003ve}.
Radiative non-leptonic kaon decays, such as
$K_{L,S}\rightarrow\pi^+\pi^-\gamma$
and $K^+\rightarrow\pi^+\pi ^0\gamma$ are
dominated by long distance contributions.
The study of these decays leads to  chiral tests and in principle to disentangle
the small short distance contribution. This small short distance  part may lead also
to interesting CP violating observables in the Standard Model (SM) and Beyond (BSM) \cite{CP}.

$K\rightarrow \pi\pi\gamma$ amplitudes   contain two types of
contributions:  inner brems\-strah\-lung (IB) and direct
emission (DE).
Due to the pole in the photon energy the IB amplitude is generally enhanced
 compared to DE; however the IB components of $K_L\rightarrow \pi^+\pi^-\gamma$
and $ K^+\rightarrow \pi^+ \pi^0\gamma$ are
suppressed due, respectively,
to
CP invariance and
 to the $\Delta I=1/2 $ rule.
Then, the  DE amplitude  of these channels, that   are the  non-trivial part of
 these decays, might
also be easier to detect.  DE contributions can be
decomposed into electric (E1,E2) and magnetic (M1,M2) ones \cite{DMS92}.
The magnetic contribution  to
$K_L\rightarrow \pi^+\pi^-\gamma$
is accurately measured \cite{KTeV,PDG06}  and  a
 clear and large photon energy  dependence has been found.

The question of the presence of the form factor, i.e.
the size of vector
meson dominance in  weak  amplitudes, is
motivated not only by phenomenological reasons, as in
$K_L\rightarrow \pi^+\pi^-\gamma$, but it has
also theoretical motivations \cite{Rafael}. For instance,
it improves the matching  of long and short
distance contributions \cite{Rafael}.

Published data for $ K^\pm\rightarrow \pi^\pm \pi^0\gamma$
  are consistent with a dominant magnetic amplitude and no evidence for E1 transitions
 \cite{PDG06}.
However, preliminary data from NA48/2 at CERN show a non-vanishing
interference, due to E1 transitions \cite{NA48/2,RaggiCundy}: the
size of this contribution will shed light on  $O(p^4)$ chiral
perturbation theory ($\chi$PT) counterterm coefficients
\cite{GI98,EKW93,DP98}. The energy
dependence of the $ K^+\rightarrow \pi^+ \pi^0\gamma$ 
magnetic amplitude has not been tested/observed yet; we think
it is important to understand if the
form factor in the magnetic contribution affects the determination of the electric contributions: in this paper 
 we  complement Ref.\cite{D'Ambrosio:2000yc} with this perspective.

Long ago  N. Christ used    a particular set
of $ K^\pm\rightarrow \pi^\pm \pi^0\gamma$  Dalitz variables: 
the photon energy and the charged pion kinetic energy in the kaon rest frame   \cite{Christ}. NA48/2 \cite{NA48/2,RaggiCundy} wants to
 perform  an analysis using these variables too; in the following we give 
a kinematical distribution in these variables accounting 
 for the 
$m_{\pi^+} -m_{\pi^0}$ corrections.
Furthermore,
the results in Ref.\cite{D'Ambrosio:2000yc}
considered essentially the central value of  branching ratio from E787 \cite{BNL00},
 $(4.7{{\pm} }0.8){{\times} }10^{-6}$.
Recent results   \cite{Aliev,RaggiCundy}
seem to prefer smaller values, $(2\sim 3)\times 10^{-6} $;
while
this is not shocking by itself, it has some impact if we include the form factor as we shall see.
We discuss   $K\rightarrow \pi\pi \gamma$
kinematics and Low theorem in Section 2;
 in  Section 3 we carry out some
theoretical-phenomenological considerations on $K\rightarrow \pi\pi \gamma$.
Numerical  results are summarized in Section 4.

\section{Kinematics and Low Theorem}
The general invariant amplitude of $K\rightarrow \pi\pi\gamma$ can be
defined as follows \cite{ENP94,DI95}
\begin{equation*}
A[K(p)\rightarrow\pi_1(p_1)\pi_2(p_2)\gamma(q,\epsilon)]=\epsilon^\mu(q)
M_\mu (q,p_1,p_2), \end{equation*}
where $\epsilon_\mu(q)$ is the photon polarization and $M_\mu$ is
decomposed into an electric $E$ and a magnetic $M$ amplitudes as
$$M_\mu={\frac{E(z_i)}{m_K^3}}[p_1{\cdot}q p_{2\mu}-p_2{\cdot}q
p_{1\mu}]+{\frac{M(z_i)}{m_K^3}} \epsilon_{\mu\nu\alpha\beta} p_1^\nu
p_2^\alpha q^\beta,$$ 
with
\begin{eqnarray}
z_i=\frac{q{\cdot}p_i}{m_K^2},\;(i=1,2),\;\; z_3=\frac{p{\cdot}q}{m_K^2}
,\;\;z_3=z_1+z_2.\hspace{0.25cm}\label{amplEM}
\end{eqnarray}
The double differential rate for the unpolarized photon is
\begin{eqnarray}
\frac{\partial^2\Gamma}{\partial z_1\partial z_2}=&\hspace{-0.6cm}\frac{\displaystyle{m_K}}{(\displaystyle{4 \pi)^3}}
\left[|E(z_i)|^2+ |M(z_i)|^2\right]\times\quad \label{widthEM}\\ &\hspace{-0.6cm} \left[ z_1 z_2(1-2 z_3-r_1^2-r_2^2)-r_1^2 z_2^2-r_2^2 z_1^2\right],\nonumber
\end{eqnarray}
where $r_i=m_{\pi_i}/m_K$. Low theorem establishes a precise relation between radiative and non-radiative
amplitudes in the limit of $E_\gamma\rightarrow 0$
\cite{DMS92,DEIN95}. Then we can  generally write the
relation between the bremsstrahlung amplitude, $E_{IB} $, in $K\to
\pi \pi \gamma $ decays and the on-shell amplitude, $ A(K\to \pi
\pi) $:
\begin{equation}
E_{IB}(z_{i})\doteq {\frac{eA(K\to \pi _{1}\pi _{2})}{m_{K}z_{3}}}\left( {%
\frac{Q_{2}}{z_{2}}}-{\frac{Q_{1}}{z_{1}}}\right) ,  \label{Ecbremss}
\end{equation}
where   $Q_{i}$ is the $\pi _i$ charge. Direct emission amplitudes
are defined by subtracting this contribution from the total amplitude. In the  $K^+$ case
\begin{eqnarray}
E_{IB}(K^{+ }) &=&e^{i\delta _{2}}\left( {\frac{3eA_{2}}{2m_{K}z_{+
}z_{3}}}\right)   \label{kppgele}
\end{eqnarray}
where
 $A(K^{+}\to \pi ^{+}\pi ^{0})={\frac{3}{2}}A_{2}e^{i\delta _{2}}$
and $z_+$ refers to the charged pion. Using the experimental value for
$B(K^{+}\rightarrow \pi ^{+}\pi ^{0})$ we obtain the branching
ratios for the Inner Bremsstrahlung shown in TAB. \ref{tab:brem}.
\begin{table}[<h>]
\begin{tabular}{c|c}
$T_{c}^{\ast }$-range in ${\rm
MeV}$
&$B(K^{+}\rightarrow \pi ^{+}\pi ^{0}\gamma)_{\rm IB}$\\ \hline
$\left[55, 90\right]$
 &$ (2.61){{\times} }10^{-4}$ \\
$\left[0, 80\right]$
 & $ (1.84){{\times} }10^{-4}$\\
\end{tabular}
{\caption{\label{tab:brem} Inner Bremsstrahlung}}
\end{table}

In $K_L\rightarrow \pi^+\pi^-\gamma$, the most common variables are:
\emph{i)} the photon energy in the kaon rest frame $E_\gamma^*$,
and  \emph{ii)} the angle $\theta$ between the photon and $\pi^+$
momenta in the di-pion rest frame. The relations between
$E_\gamma^*$, $\theta$ and the $z_i$ are:
\begin{eqnarray}
z_3=\frac{E_\gamma^*}{m_K},\;\;\;\;z_{\pm}=\frac{E_\gamma^*}{2m_K}(1\mp\beta {\rm cos}
\theta),
\end{eqnarray}
where $\beta=\sqrt{1-4m_\pi^2/(m_K^2-2m_K E_\gamma^*)}$.
Then the differential rate is
\begin{eqnarray}
\frac{\partial^2 \Gamma}{\partial E_\gamma^*\partial
{\rm  cos}\theta}=\frac{(E_\gamma^*)^3\beta^3}
{512\pi^3m_K^3}\left(1-\frac{2E_\gamma^*}{m_K}\right){\rm sin}^2\theta\nonumber\\(|E|^2+|M|^2).
\end{eqnarray}

Since  three photons will be detected in the
$K^+\rightarrow\pi^+\pi^0\gamma$ measurement, it is very useful to study
the differential rate as a function of: \emph{i)} the charged pion
kinetic energy in the $K^+$ rest frame $T_c^*$, and \emph{ii)}
$W^{2}=(q{{\cdot} }p_{K})(q\cdot p_{+})/(m_{\pi ^{+}}^{2}m_{K}^{2})$
\cite{DMS92}. These two variables are related to the $z_{i}$ by
\begin{eqnarray}
\hspace{-0.6cm} z_{0}=&\Frac{m_{K}^{2}+m_{\pi
^{+}}^{2}-m_{\pi^{0}}^{2}-2m_{K}m_{\pi ^{+}}-2m_{K}T_{c}^{\ast }}{\displaystyle 2m_{K}^{2}},
\nonumber  \\
z_{3}z_{+}=&\hspace{-5.3cm}\frac{m_{\pi ^{+}}^{2}}{m_{K}^{2}}W^{2}.\label{Z0TC}
\end{eqnarray}
The advantage of  these variables is that, under the assumption of constant $E_{DE}$ and $M_{DE}$, one can easily disentangle the different contributions of
the IB, DE
amplitudes, and interference term between IB and DE
\begin{eqnarray}
\Frac{\partial ^{2}\Gamma }{\partial T_{c}^{\ast }\partial W^{2}}
&=\Frac{\partial ^{2}\Gamma _{IB}}{\partial T_{c}^{\ast }\partial
W^{2}}\left[1+\Frac{m_{\pi ^{+}}^{2}}{m_{K}}2Re
\left(\Frac{E_{DE}}{e{\cal A}}\right)W^{2}
\right. \nonumber \\
&\left. +\Frac{m_{\pi^{+}}^{4}}{m_{K}^{2}}\left(\left|\Frac{E_{DE}}{e{\cal A}}\right|^{2}+\left|\Frac{M_{DE}}{
e{\cal A}}\right|^{2}\right)W^{4}\right], \qquad \label{dWEM}
\end{eqnarray}
where 
\beq
{\cal A}=A(K^{+}\rightarrow \pi ^{+}\pi ^{0}).\label{eq:calAkpp}\eeq

\subsection{Kinematics in  Christ's variables $T_c^*$ and
$E_\gamma^*$}

Motivated by the upcoming NA48 measurements we have also studied
 the Dalitz plot distributions in  the kinematical variables
$T_c^*$
 and the 
photon energy in the kaon rest frame,
$E_\gamma^*$.
In this way
 we extend the work by N. Christ
\cite{Christ} by including the terms proportional to
$m_{\pi ^+}-m_{\pi ^0}$; in these variables
  the
double differential rate for $K^+\rightarrow \pi^+\pi^0\gamma$ is written as
\begin{eqnarray}
\frac{\partial ^{2}\Gamma }{\partial T_{c}^{\ast }\partial
E_{\gamma }^{\ast }} =
\frac{\partial ^{2}\Gamma _{IB}}{\partial
T_{c}^{\ast }\partial E_{\gamma }^{\ast }}
\left[
\hspace{0.3cm}
 1
\hspace{0.7cm}
+\right.
&&\label{eq:TcEg}\\
&&\hspace{-4cm} 2Re\left(
\frac{E_{DE}}{e{\cal A}
}\right) \left( \frac{m_{K}}{2}-E_{0}-\frac{\delta
\mu ^{2}}{2m_{K}}\right)\frac{E_{\gamma
}^{\ast }}{m_{K}}+\left.  \right. \nonumber \\
&&
\hspace{-5.2cm}
\left. \left( \left| \frac{E_{DE}}{e{\cal A}}\right| ^{2}+\left| \frac{M_{DE}}{
e{\cal A}
}\right| ^{2}\right) \left(\frac{m_{K}}{2}-E_{0}-\frac{\delta \mu ^{2}}{2m_{K}}%
\right)^{2}\frac{E_{\gamma }^{\ast 2}}{m_{K}^{2}}\right] , \nonumber
\end{eqnarray}
where $E_{0}$ is the $\pi ^0$-energy,
$$\delta \mu ^{2} =m_{\pi ^+}^{2}-m_{\pi ^0}^{2}$$
and  the Inner Bremsstrahlung
differential rate    is written in terms of angle, $\theta $, between the
$\pi ^+$ and $\gamma$
momenta in the kaon rest frame
\begin{eqnarray*}
\hspace{-0.cm}\cos \theta &=&\\
&&\hspace{-1.cm} \Frac{
  m_{K}^{2}-2 \, m_{K}E_{\gamma }^{\ast
}+ 2\,(T_{c}^{\ast
}+m_{\pi ^+})(E_{\gamma }^{\ast }-m_K)+\delta \mu ^{2}}
{2 \,  p_{\pi^+}^{\ast } \, E_{\gamma }^{\ast }} \nonumber
\end{eqnarray*}
\begin{equation*}
p_{\pi ^+}^{\ast } =\sqrt{T_{c}^{\ast }(T_{c}^{\ast }+2 m_{\pi ^+})} ,
\end{equation*}
\[
\frac{\partial ^{2}\Gamma _{IB}}{\partial T_{c}^{\ast }\partial
E_{\gamma
}^{\ast }}=\frac{\alpha }{\left( 4\pi \right) ^{2}}\frac{{p_{+}^{\ast }}^2}{%
m_{K}^{3}}\sin ^{2}\theta \frac{\left|   {\cal A}\right| ^{2}
}{\left(\displaystyle\frac{m_{K}}{2}-E_{0}-\displaystyle\frac{\delta
\mu ^{2}}{2m_{K}}\right)^{2}}.
\]
It is possible to pass from eq.(\ref{eq:TcEg}) to  eq.(\ref{dWEM}) through
\[
W^{2}=\frac{E_{\gamma }^{\ast }}{m_{\pi ^+}^2}(E_{\gamma }^{\ast
}+T_{c}^{\ast }+m_{\pi ^+}-\frac{m_{K}}{2}-\frac{\delta \mu
^{2}}{2m_{K}}).
\]

A contour plot of the IB amplitude in the Dalitz plot of Christ's
variables $T_c^*$ and $E_{\gamma}^*$ is shown in FIG.1. After
integrating eq.(\ref{eq:TcEg}) on $E_\gamma^*$, one obtains
($E_{c}^{\ast }=T_{c}^{\ast }+m_{\pi^+}$) \noindent
\begin{eqnarray*}
\frac{d\Gamma}{d T_{c}^{\ast}}& \hspace{-1.3cm}=\Frac{\alpha
\left|
{\cal A}\right|^2
}{(2\pi)^2}\left\{\Frac{2
\left[\Frac{E_{c}^{\ast
}}{m_K}\log
( \Frac{E_{c}^{\ast }+p_{+}^{\ast
}}{m_{\pi^+}})
-\Frac{p_{+}^{\ast }}{m_K}\right]
}{m_K\left(m_K-2E_{c}^{\ast
}+\displaystyle\frac{\delta\mu^2}{m_K}\right)}\right.\\
 &\hspace{-1cm}\left. +Re\left( \Frac{E_{DE}}{e{\cal A}}\right)\Frac{m_K-2
E_{c}^{\ast }+\displaystyle\frac{\delta \mu^2}{m_K}}{2m_K^2 }
   \left[\Frac{1}{2}
   \left(1-\Frac{m_{\pi^+}^2}{m_K^2}\right) \cdot \right.\right.\,
\nonumber\\
 &\hspace{-1.3cm}\left.\left. \log
   \left(\Frac{m_K-E_{c}^{\ast }+p_{+}^{\ast }}
   {m_K-E_{c}^{\ast }-p_{+}^{\ast }}\right)-\Frac{m_{\pi^+}^2}{m_K^2}
   \log \left(\Frac{E_{c}^{\ast }+p_{+}^{\ast }}{m_{\pi^+}}\right)
   -\Frac{p_{+}^{\ast }}{m_K}\right]\right.
\nonumber\\
&\left. \hspace{-1.5cm}+\left( \left| \Frac{E_{DE}}{e{\cal A}}\right| ^{2}+\left| \Frac{M_{DE}}{e{\cal A}
}\right| ^{2}\right)\Frac{{p_{+}^{\ast }}^3 \left(m_K-2
E_{c}^{\ast }+\displaystyle\Frac{\delta \mu ^2}{m_K}\right)^3}{24 m_K^2
   \left((m_K-E_{c}^{\ast })^2-{p_{+}^{\ast }}^2\right)^2}\right\}
\end{eqnarray*}

\begin{figure}[t]
\centering
\epsfysize=6cm\epsfxsize=7.5cm\epsfbox{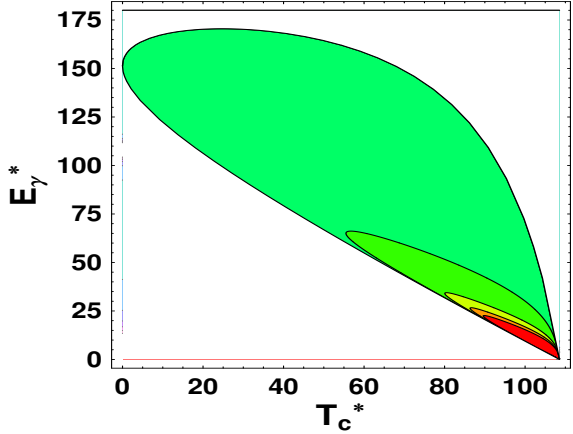}
\parbox{7.5cm}{\caption{ Dalitz plot of Christ's variables $T_c^\ast$ and $E_{\gamma}^\ast$
and contour plot of the 
\red{ Inner Bremsstrahlung} 
amplitude: 
the right corner (in red) has a greater  density.
}}\label{fig:DalitzIBChrist}
\end{figure}
\vskip0.3cm
\section{Experimental status and theoretical predictions}

Electric and magnetic dipole (E1 and M1) transitions appear already at $\mathcal{O}(p^{4})$
\cite{EKW93,DP98}.
Since  $K^+$ is not a CP eigenstate,
 M1 and E1 transitions are CP-allowed  in $K^+$,  while
for  $K_L$, CP-symmetry allows M1 transitions and inhibits E1 transitions.
The $\mathcal{O}(p^{4})$ contributions to $K^+$ electric transitions
 can be  parametrized as
\bea
E_{1}(K^{+ }) &=&-e^{i\delta _{c}}{\frac{eG_{8}m_{K}^{3}}{8\pi ^{2}F_{\pi }}}N_{E_{1}}^{(4)},\label{Ec}
\eea
where
$G_{8} = 9\cdot 10^{-6}$ GeV$^{-2}$ is obtained from
the $\Delta I=1/2$ contribution to $K_S\to \pi \pi $ at $\mathcal{O}%
(p^{2})$ and
  $N_{E_{1}}^{(4)}$ is the relevant $\mathcal{O}(p^{4})$ weak counterterm combination 
\cite{EKW93,DI95,DP98}
\beqa
N_{E_1}^{(4)} &=& (4 \pi)^2 \left[ N_{14}-N_{15}-N_{16}-N_{17}\right] \nonumber
\\
&  \stackrel{FM}{=} &
-  k_f  \Frac{ 8 \pi^2 F_\pi^2} {m_V^2}
 = -(0.4 \div 1).\label{eq:fac}\eeqa
The second line is the theoretical prediction based on the factorization model (FM) \cite{EKW93,DI95,DP98,GI98,Bruno:1992za} parametrized by
the a coefficient $k_f$.
From na\"{i}ve dimensional analysis (NDA)  we expect $N_{E_{1}}^{(4)}$ of order one;
in fact this value is  expected  from VMD and factorization \cite{GI98,DP98}. 
The sign in (\ref{Ec}) and (\ref{eq:fac}) leads to a constructive interference among $E_{1}$ and $%
E_{IB}$ \cite{DI95} but
{\it so far there is no experimental evidence  of such interference}, up to the
new NA48 result \cite{RaggiCundy}, as we shall see.

Actually, the present  bounds \cite{BNL00,Aliev} are very close to
the theoretical predictions \cite{GI98,EKW93,DP98}; in TAB. \ref{tab:kppg} we report instead the results
for a  destructive interference, $N_{E_{1}}^{(4)}=0.4$,  as reported by  NA48 \cite{RaggiCundy}. The
interference term scales linearly with $N_{E_{1}}^{(4)}$, then
branchings for different  $N_{E_{1}}^{(4)}$  values  are easily
obtained.
\noindent The same counterterm coefficient, $N_{E_{1}}^{(4)}$,
appears in the DE component of $K_S\to \pi ^+ \pi ^- \gamma$ \cite{DI95,DMS93},
 however the present experimental bound  \cite{Taureg:1976mh} is not at the level to
compete with
 the one from $K^+\to \pi ^+ \pi ^0 \gamma$.

No $E1$ (CP violating) transitions for $K_L$ have been observed yet
\cite{KTeV}; while the magnetic contributions, $M1$ is responsible for the observed
   ${\rm B}(K_L\to \pi ^{+}\pi ^{-}\gamma)_{DE}$.  
The leading order magnetic amplitude, $M(z_i)$, to eq. (\ref{amplEM}),
starts at $O(p^4)$ \cite{ENP94,DEIN95}
\begin{eqnarray}
M^{(4)}_{L}=\frac{e G_8 m_K^3}{2\pi^2 F}(a_2+2a_4),\label{ML4}\\
M^{(4)}_{+}=-\frac{e G_8 m_K^3}{4\pi^2 F}[2+3(2 a_3-a_2)].\label{MP4}
\end{eqnarray}

The subscripts $L$ and $+$ denote  $K_L\rightarrow\pi^+\pi^-\gamma$ and
$K^+\rightarrow\pi^+\pi^0\gamma$ respectively.
The $a_i$'s parts of the
above amplitudes come from the local weak lagrangian ${\cal
L}_4^{\Delta S=1}$  and are expected of order one \cite{Bijnens:1992ky,ENP94}; these are also called direct contributions.
They were originally derived using the factorized form,
current-current, of the  weak chiral lagrangian with one current ${\cal O}(p^3)$ from the WZW
 lagrangian and the other  current from  the usual ${\cal O}(p)$  current \cite{Bijnens:1992ky}.
Later it was found
 that also vectors and axials (VMD) were  contributing to these  coefficients ($a_i$'s)   
 \cite{DP98}.
  The factor \lq \lq 2\rq\rq in (\ref{MP4}) is the
contribution from the $K^+\rightarrow \lq\lq \pi ^+ $\rq\rq
$\rightarrow\pi^+\pi^0\gamma$, where we have  first  a pure weak
transition and then a  Wess-Zumino-Witten one
and thus it is completely predicted (indirect contribution). There are too many
$a_i$'s to be fixed  phenomenologically
and possibly large
${\cal O}(p^6)$ corrections to $K_L\rightarrow\pi^+\pi^-\gamma$ in (\ref{ML4})
from $\eta '$-exchange.

An interesting progress to the  understanding of these decays has been driven by KTeV;
in order to fit the $K_L\rightarrow\pi^+\pi^-\gamma$ data showing 
a clear  photon energy dependence in the magnetic term,
$M(z_i)$, KTeV has used a linear fit, a quadratic fit or a pole fit of  the form
\begin{equation}
{ M_L}= {e\ |h_M|}\left( \Frac{b}{1-\frac{m_K^2}{m_V^2}+\frac{2m_K}{m_V^2}
{E_\gamma^*}}+1\right).
\label{EVMD}
\end{equation}
 Interestingly, data prefer this pole fit
to linear and quadratic fit \cite{KTeV}. The rate and the photon energy spectrum 
fix   $|h_M| =(9.4\pm 0.8)\cdot 10^{-7} $ and $b=-1.243\pm 0.033$  \cite{KTeV}.  This
phenomenological description has been comforted by  NA48
\cite{Lai:2003ad} and KTeV  \cite{Abouzaid:2005te}  in the channel
$K_L\to \pi ^{+}\pi ^{-} e^+ e^-$. 
Comparison of (\ref{EVMD}) with  (\ref{ML4}) leads to the following theoretical consequences:
i) the value of $h_M$ fixes the size of the  $a_i$'s: $\sim {\cal{O}} (1)$ and ii)
the presence of a relatively large component, $b$, with form factor tells us that  
 VMD plays a major r\^ole in
these coefficients. 
The usual vector formulation, \lq \lq $V^\mu $\rq\rq, is very successful to this description: there
are vector and axial contributions to the  $a_i$'s \cite{DP98}.
However
 the tensor formulation of the 
vectors, \lq \lq $V^{\mu \nu} $\rq\rq,  very successful in the strong sector, does not have any 
vector contributions to the  $a_i$'s \cite{EKW93,D'Ambrosio:2000yc}.
 The lack of vector contributions  to the  $a_i$'s and possible large ${\cal O}(p^6)$ contribution  
to (\ref{ML4})
from $\eta '$-exchange does not explain the 
 observed large form factor 
in (\ref{EVMD}) and puts tension in the chiral expansion with large ${\cal O}(p^6)$ and small ${\cal O}(p^4)$ 
contributions \cite{ENP94}. 
Actually, in  Ref. \cite{Ecker:1990in} it was already  shown that 
while  
the odd-parity couplings 
 to $V \to P\gamma $ decays, relevant  to the anomalous $K\to \pi \pi\gamma $ decays, had the proper
QCD behaviour  if we use the usual vector formulation, \lq \lq $V^\mu $\rq\rq,
this was not the case if we use the tensor formulation of the 
vectors, \lq \lq $V^{\mu \nu} $\rq\rq.
 As a result we believe in the large VMD contribution to the   $a_i$'s.
\begin{table}
Experimental status
\vskip0.2cm
\begin{tabular}{c|c}
REFS.
 & $B(K^{+}\rightarrow \pi ^{+}\pi ^{0}\gamma)_{\rm DE}$\\ 
&$T_{c}^{\ast }\in \left[55, 90\right]$ ${\rm}$\\ \hline \hline
BNL E787   \cite{BNL00} &$(4.7{{\pm} }0.9){{\times} }10^{-6}$\\ \hline
PDG 06   \cite{PDG06} &$(4.4{{\pm} }0.7){{\times} }10^{-6}$\\ \hline \hline
KEK-E470   \cite{Aliev}& $(3.8{{\pm} }0.8\pm 0.7){{\times} }10^{-6}$ \\
NA48/2  \cite{RaggiCundy}
& $(2.22\pm0.13\pm0.05){{\times} }10^{-6}$ \\ \hline\hline 
\multicolumn{2}{c}{}\\
{NA48/2  analysis \cite{RaggiCundy}}&$T_{c}^{\ast }\in \left[0, 80\right]$ MeV\\  \hline
$B(K^{+}\rightarrow \pi ^{+}\pi ^{0}\gamma)^{\rm INT}$& $(-4.91\pm 2.00){{\times} }10^{-6}$\\\hline
$B(K^{+}\rightarrow \pi ^{+}\pi ^{0}\gamma)^{\rm DE}$&$(6.16\pm0.79){{\times} }10^{-6}$\\\hline\hline
\multicolumn{2}{c}{}\\\hline\hline
\multicolumn{2}{c}{Theory predictions}\\
$T_{c}^{\ast }$-range in ${\rm MeV}$ &$B(K^{+}\rightarrow \pi
^{+}\pi ^{0}\gamma)_{\rm INT}^{^{{N_{E_1}^{(4)}}=0.4}}$\\ \hline
$\left[55, 90\right]$
 & $ -(3.52){{\times} }10^{-6}$\\
$\left[0, 80\right]$
 & $ -(4.70){{\times} }10^{-6}$\\
\multicolumn{2}{c}{}\\
$T_{c}^{\ast }$-range in ${\rm
MeV}$
&$B(K^{+}\rightarrow \pi ^{+}\pi ^{0}\gamma)_{\rm DE}^{^{{a_i}=0}}$\\ \hline
$\left[55, 90\right]$
 & $ (3.55){{\times} }10^{-6}$\\
$\left[0, 80\right]$
 & $ (6.57){{\times} }10^{-6}$\\
\end{tabular}
\vskip0.3cm
{\caption{\label{tab:kppg} Experimental and theoretical status\\
The table shows 
 in the first two lines   the PDG 06 value along with its most precise measurement, BNL  E787  and in the next
 two lines two subsequent measurements, KEK-E470 and NA48/2, pointing towards  a 
smaller value of the branching ratio. All these values are obtained with vanishing interference. 
To compare  NA48/2 with other experiments we extrapolated the NA48/2 value, obtained in the kinematical range 
$0 \leq T_{c}^{\ast }\leq 80\ {\rm MeV}$, to the kinematical range 
$55 \leq T_{c}^{\ast }\leq 90\ {\rm MeV}$, assuming a 
constant magnetic term.  
We report  in the fifth and sixth row,  the  interference and the direct
 emission contributions  determined simultaneously by NA48/2 \cite{{RaggiCundy}}. 
 The INT theoretical branching ratio  with a value  
of the 
weak counterterm 
combination in eq. (\ref{eq:fac})   so to match the NA48/2 result, $ N_{E_1}^{(4)}=0.4$, is then shown.
For comparison the DE  theoretical branching ratios obtained   with $a_i=0$ are reported in the last rows.}}
\end{table}

We show in TAB. \ref{tab:kppg} an updated  $K^+\rightarrow\pi^+\pi^0\gamma$ experimental and theoretical 
situation. 
We write 
in the first two lines  the PDG 06 value along with its most precise measurement, BNL  E787; in the next
 two lines two subsequent measurements, KEK-E470 and NA48/2, that as  we can see, point towards  a 
smaller value of the branching ratio. All these values are obtained with vanishing interference. Also 
to compare  NA48/2 with other experiments we extrapolated the NA48/2 value, obtained in the kinematical range 
$0 \leq T_{c}^{\ast }\leq 80\ {\rm MeV}$, to the kinematical range $55 \leq T_{c}^{\ast }\leq 90\ {\rm MeV}$ assuming a 
constant magnetic term.  
Interestingly NA48/2 has also done the analysis to determine simultaneously both the  interference and the direct
 emission contributions \cite{{RaggiCundy}}; 
we show in  TAB. \ref{tab:kppg} this NA48/2  analysis 
 showing non-vanishing  values for both the  interference and the direct
 emission contributions.

We  also show  in  TAB. \ref{tab:kppg} for comparison some theoretical predictions  for the interference term
and the direct emission term. 
Regarding the interference term 
we use  a specific value of  the 
weak counterterm 
combination in eq. (\ref{eq:fac}): $ N_{E_1}^{(4)}=0.4$,   so to match the NA48/2 result. 
Regarding the DE contribution we take as comparison the $a_i=0$.

 Following Ref. \cite{D'Ambrosio:2000yc} a  correlated
analysis of the $K^+\rightarrow\pi^+\pi^0\gamma$ and $K_L\to \pi
^{+}\pi ^{-}\gamma$ decays can be performed.   VMD and phenomenology,
i.e.  eq. (\ref{EVMD}),   imply the following decomposition:
\begin{eqnarray}
M_+ (z_i)=M^{\rm pole}_+ +M^{\rm const.}_+\label{MVMD}
\end{eqnarray}
It is interesting that all the weak couplings involving vectors, and
consequently $M_+^{\rm }$ and  $M_L^{\rm VMD}$,
 may be written in terms of only
one coupling, $\eta _V$ \cite{DP98}. Thus  we parametrize all our ignorance in  $ M^+_{\rm pole}$ with $\eta _V$:
\begin{eqnarray}
M_+^{\rm pole}(z_i)=-\frac{e G_8 m_K^3}{4\pi^2
F}r_V \left(\Frac{1-2z_3+
\frac{m_V^2}{m_K^2}\eta_V}{1-\frac{m_K^2}{m_V^2}+\frac{2m_K^2}{m_V^2}z_3}+\right.\nonumber\\  \left.
 \Frac{2z_+-\frac{m_V^2}{2m_K^2}\eta_V}{1-\frac{2m_K^2}{m_V^2}z_+}
+\Frac{2z_0+\frac{m_V^2}{m_K^2}\eta_V}{1-\frac{2m_K^2}{m_V^2}z_0}\right),\hspace{0.4cm}
\label{KPVMD}
\end{eqnarray}
where
\begin{eqnarray}
r_V=\frac{32\sqrt{2}\pi^2 f_V h_V m_K^2}{3 m_V^2},
\end{eqnarray}
is determined  by the VMD couplings \cite{ENP94,DP98} $f_V=0.2$ and
$h_V=0.037$,  well known phenomenologically.
The rest is written as a constant  contribution,
\begin{equation}
M_+^{\rm const.}=-\frac{e G_8 m_K^3}{4\pi^2  F}A^+,
\end{equation}
where the parameter 
\begin{equation*}
A^+=2+3(2a_3-a_2)_{\rm non-VMD}+~{\rm other~~ contributions}.
\end{equation*}
must be determined phenomenologically.
 $(2a_3-a_2)_{\rm non-VMD}$ is a non-VMD contribution to $M^+$;  thus
the experimental $B(K^+\rightarrow\pi^+\pi^0\gamma)_{DE}$ can be
obtained by varying the two unknown constants $(A^+,\eta _V)$. In
FIG. 2, we vary respectively
$1\sigma$, $2\sigma$ and $3\sigma$ deviations around the BNL E787 result \cite{BNL00}
 similar to the PDG result
  in TAB. \ref{tab:kppg}. 
So we can account properly the
sensitivity of the $W-$ and
 $T_c^*-$spectrum  from these phenomenological parameters which  we will discuss  in the next
paragraph.
\begin{figure}[ht]
\centering
\epsfysize=6.8cm\epsfxsize=7.2cm\epsfbox{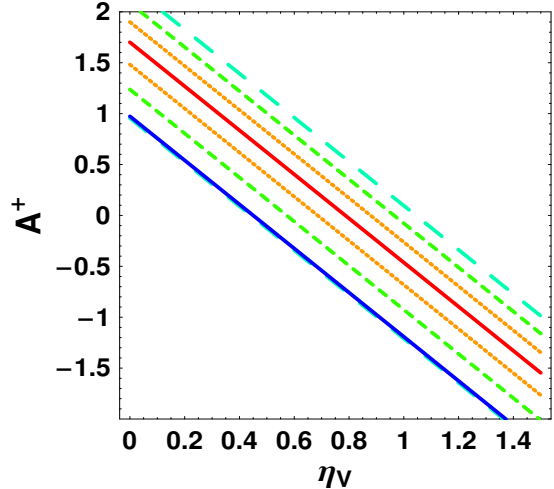}
\parbox{7.3cm}{\caption{Values of  $A^+$ and $\eta_V$  on the 
\red{ central solid line} 
generate the \red{E787} experimental value of the
branching ratio
in TAB. \ref{tab:kppg} while 
  the other lines are the
borders of strips corresponding to{\color{arancia}{ $1\sigma$}}, {\color{verde}{$2\sigma$}} and
{\color{bluazz}{$3\sigma$}} deviation from the 
E787
central value in TAB. \ref{tab:kppg}.
As a result
the {\color{blue}{lower  solid line}} corresponds to the central 
 {\color{blue} NA48} value in TAB. \ref{tab:kppg}.
 For  $\eta_V=0$ and  $A^+=2$,  the amplitude is
dominated by the WZW pole.
 }}\label{fig:etaA}
\vskip.5cm
\end{figure}
 In fact,
 recent data, shown in TAB. \ref{tab:kppg}, point towards smaller values of the branching.

\noindent NA48, in their preliminary analysis \cite{RaggiCundy}, are
able to study this decay in a new kinematical region: the charged
pion kinetic energy ranges from 0 to $80 \ {\rm MeV}$; as we can see
from FIG. 1, this region is more sensitive to the DE component.
Furthermore assuming a {\bf constant $M(z_i)$} in (\ref{amplEM})
 they
find a non-vanishing $E1$-contribution.

This result is very interesting; in particular though the size is comparable to the theoretical expectations 
in (\ref{eq:fac}) and   TAB. \ref{tab:kppg} \cite{GI98,EKW93,DP98},
the sign is opposite  as discussed before. 
In the next section we investigate
the consequences of the presence of the form factor on this NA48
measurement.

\section{Effect of the Form Factor on the Direct Emission amplitude}
The target of the following numerical studies is to understand how
to distinguish a constant magnetic term, practically  $\eta _V=0$,  from 
 the one with the form
factor in eqs. (\ref{MVMD}) and (\ref{KPVMD}), 
in  FIG. 2.    
A related question is, as we shall see,
if the possible presence of a form factor affects the determination of the interference term or even if 
the presence of the form factor could mimic the experimental evidence 
of the interference term. It is important, we think, to quantify this effect. 

Previous experiments were able to study  only
the kinematical range $T_{c}^{\ast }\in \left[55, 90\right]$ MeV \cite{BNL00,Aliev}; 
recentely NA48/2 has been able to uncover almost the full range 
$T_{c}^{\ast }\in \left[0, 80\right]$ MeV \cite{RaggiCundy}. The region with  small
 values for $T_{c}^{\ast }$ is more sensitive to DE transitions. Indeed the explicit 
IB Dalitz plot contour in  the  $T_{c}^{\ast } $ and $W$ variables   generates 
a plot similar to FIG. 1. Due to the sensitivity of the kinematical distributions 
from the size of  DE branchings we will discuss both possibilities ${\bf B_{ exp}(E787)}$ and  ${\bf B_{exp} (NA48)}$
in both kinematical regions. 
\begin{figure}[ht]
\vskip.25cm \centering
\epsfysize=5cm\epsfxsize=7.2cm\epsfbox{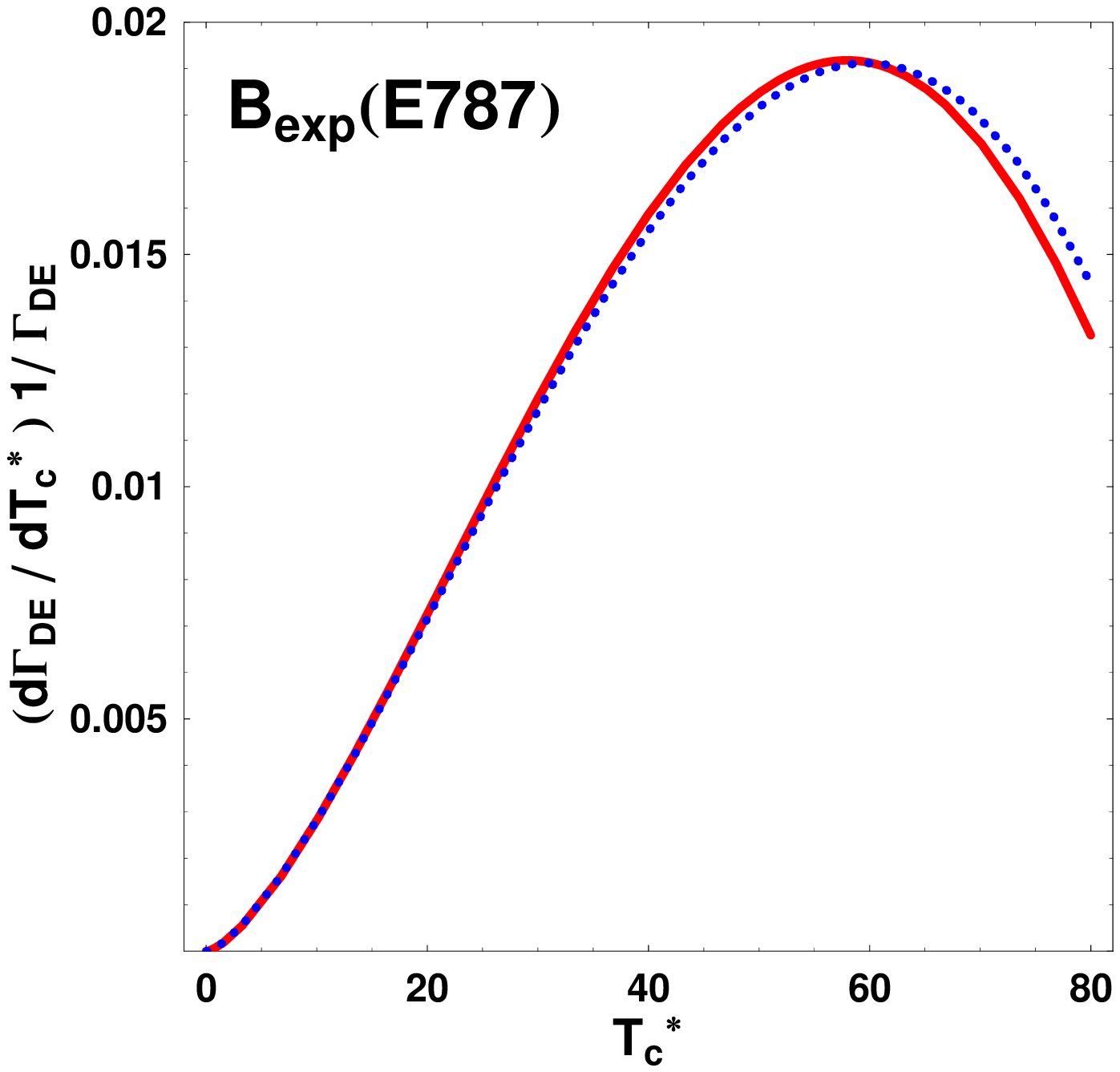}
\epsfysize=5cm\epsfxsize=7.2cm\epsfbox{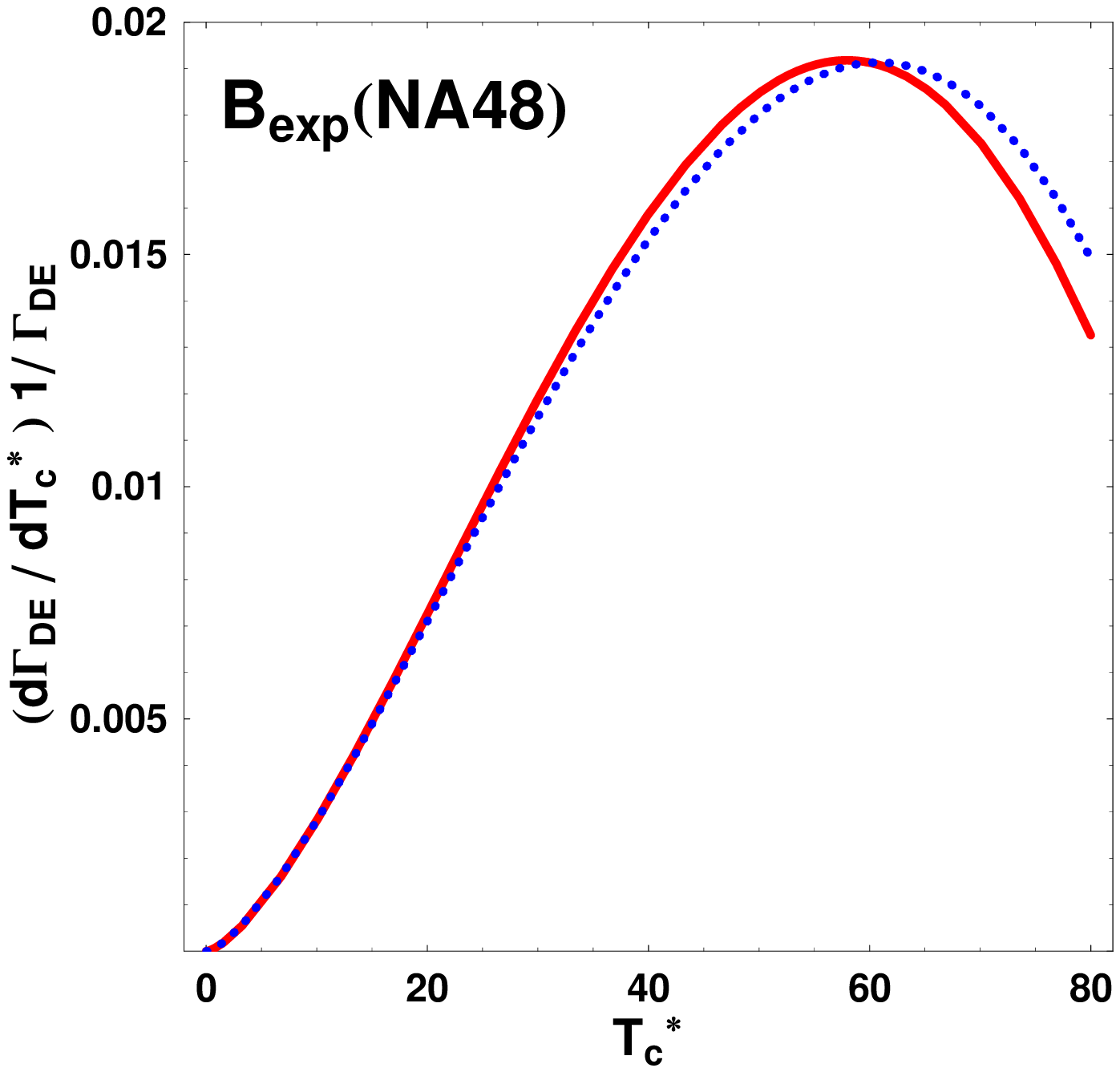}
\smallskip
\parbox{7.3cm}{\caption{Normalized $T_c^*$-spectra ($T_{c}^{\ast }\in \left[0, 80\right]$ MeV) for the DE magnetic
amplitude with   the E787  branching ratio  \cite{BNL00} in TAB. \ref{tab:kppg}
(upper plot)   and the  
 NA48/2 branching ratio in TAB. \ref{tab:kppg}
(lower plot). The \red{solid} curves correspond to a \red{constant} amplitude, the
{\color{blue} dotted} curves to a magnetic form factor with {\color{blue} $\eta_V=1.5$} and
corresponding value of $A^+$ on the central (lower) full  line of FIG. 2 for the upper (lower) plot .
}}\label{fig:TcNA48}\vskip.25cm
\end{figure}

In FIG. 3 we show
the normalized $T_c^*$-spectra  for the DE magnetic
amplitude with   the E787  branching ratio  \cite{BNL00} in
TAB. \ref{tab:kppg}
(upper plot)   and the  
 NA48/2 branching ratio in 
TAB. \ref{tab:kppg}
(lower plot). The solid curves correspond to a constant amplitude, the
 dotted curves to a magnetic form factor with $\eta_V=1.5$ and
corresponding value of $A^+$ on the central (lower) full  line of FIG. 2 for the upper (lower) plot.

Then we plot  the $W$-spectra  with 
 $T_{c}^{\ast }\in \left[55,
90\right]$ MeV in FIG. 4 and  $T_{c}^{\ast }\in \left[0,
80\right]$ MeV  in FIG. 5.
In each case we consider  
 form factors with
$\eta_V=0.5$ (dashed   curves) and $\eta_V=1.5$ (dotted  curves); the corresponding values of  
$A^+$ are determined  from FIG.2. 
\begin{figure}[hb]
\vskip.25cm\centering
\epsfysize=6cm\epsfxsize=7cm\epsfbox{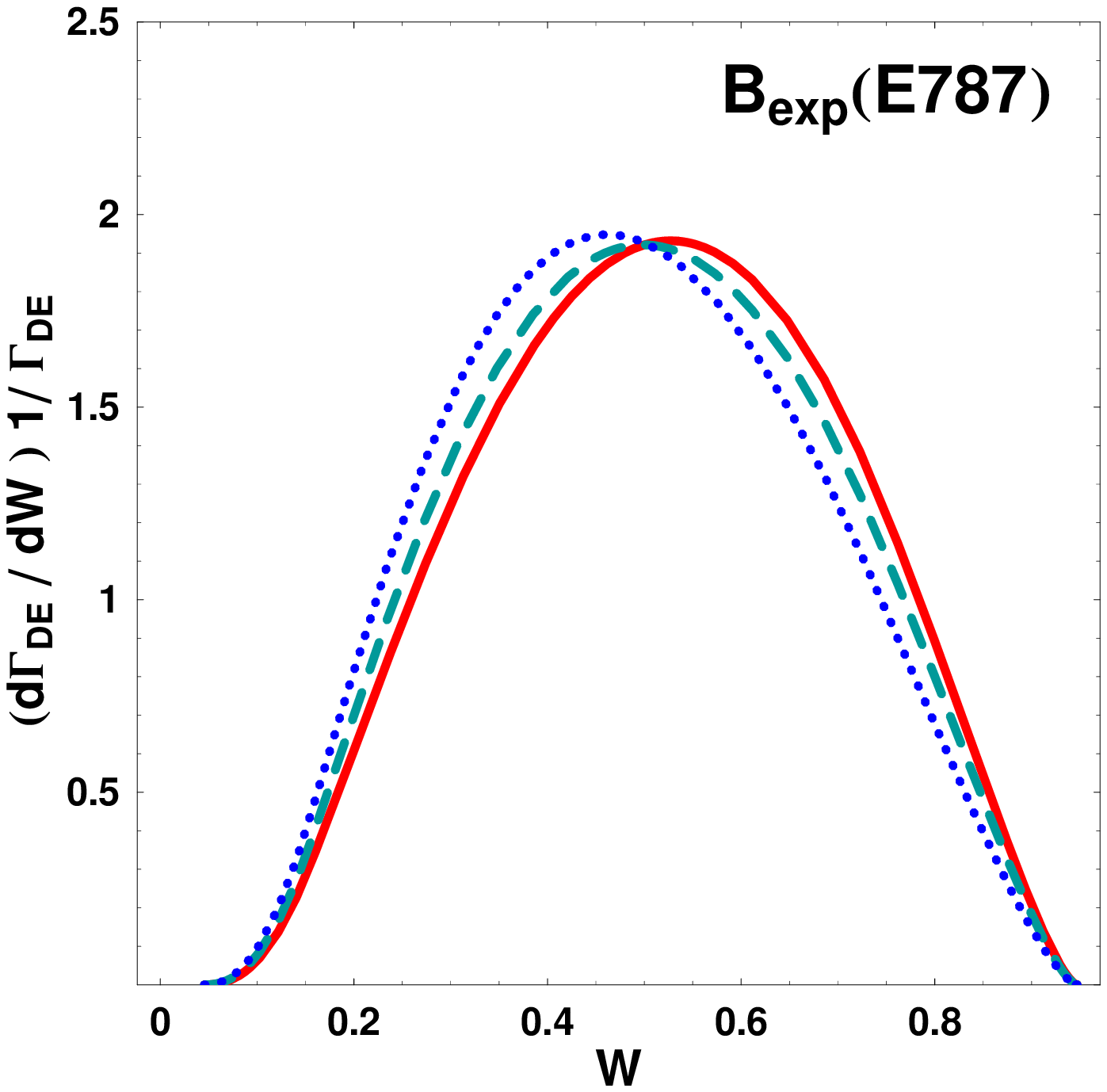}
\epsfysize=6cm\epsfxsize=7cm\epsfbox{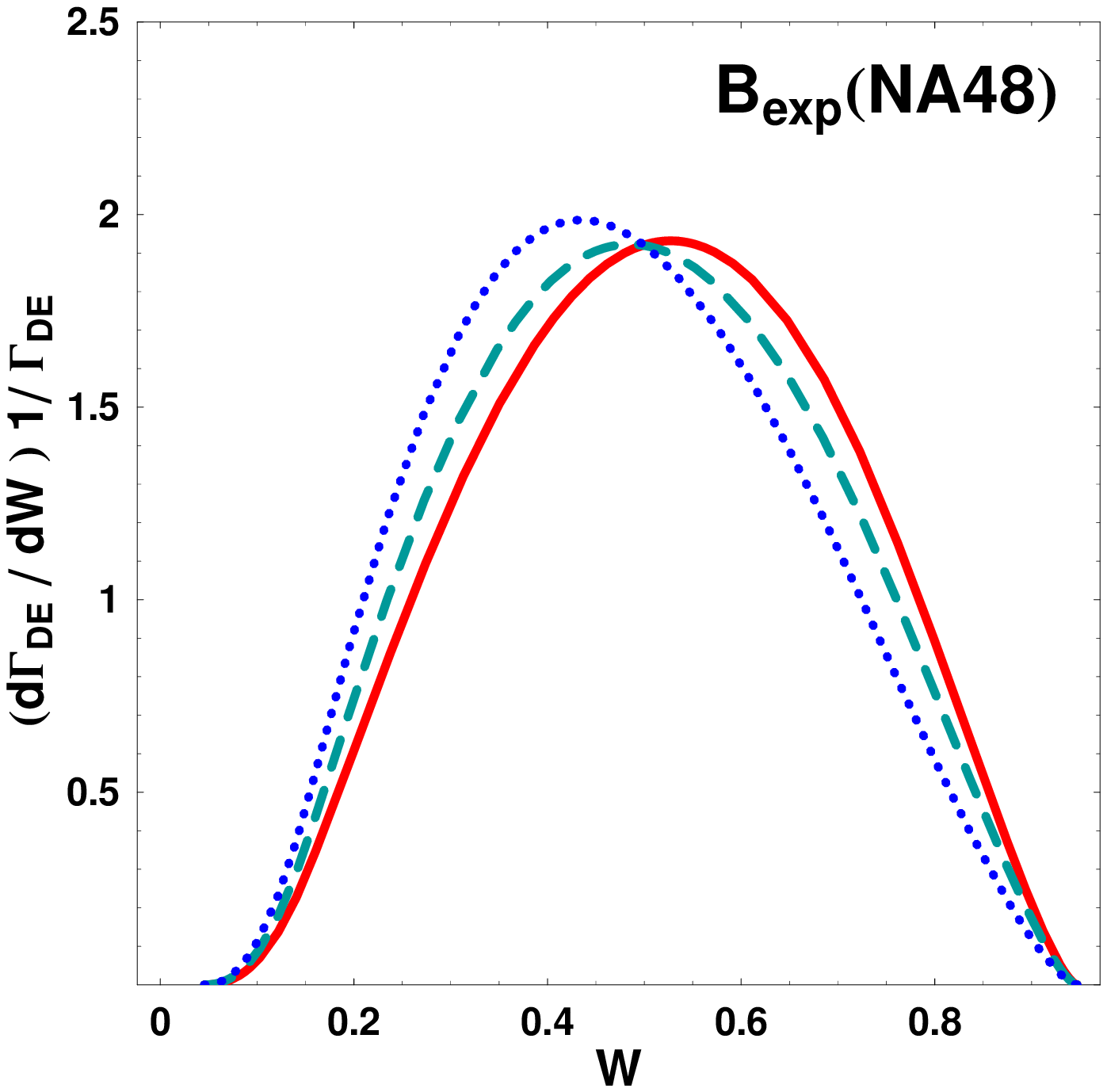}
\smallskip
\parbox{7cm}{\caption{Normalized $W$-spectra ($T_{c}^{\ast }\in \left[55, 90\right]$ MeV) for the DE magnetic
amplitude.
The upper and lower figures correspond  to the  E787 branching ratio \cite{BNL00} in 
TAB. \ref{tab:kppg}
(the central solid  line of FIG. 2)  
and to the  NA48/2 branching ratio in 
TAB. \ref{tab:kppg}
(the lower solid  line of FIG. 2) respectively.
The solid curves corresponds to a \red{constant} amplitude, while
the {\color{verdel} dashed} and {\color{blue} dotted}  curves correspond to form factors with
{\color{verdel}  $\eta_V=0.5$} and {\color{blue} $\eta_V=1.5$} respectively.
}}\label{fig:Wspect5590}
\vskip.25cm
\end{figure}

\begin{figure}[hb]
\vskip.25cm\centering
\epsfysize=6cm\epsfxsize=7cm\epsfbox{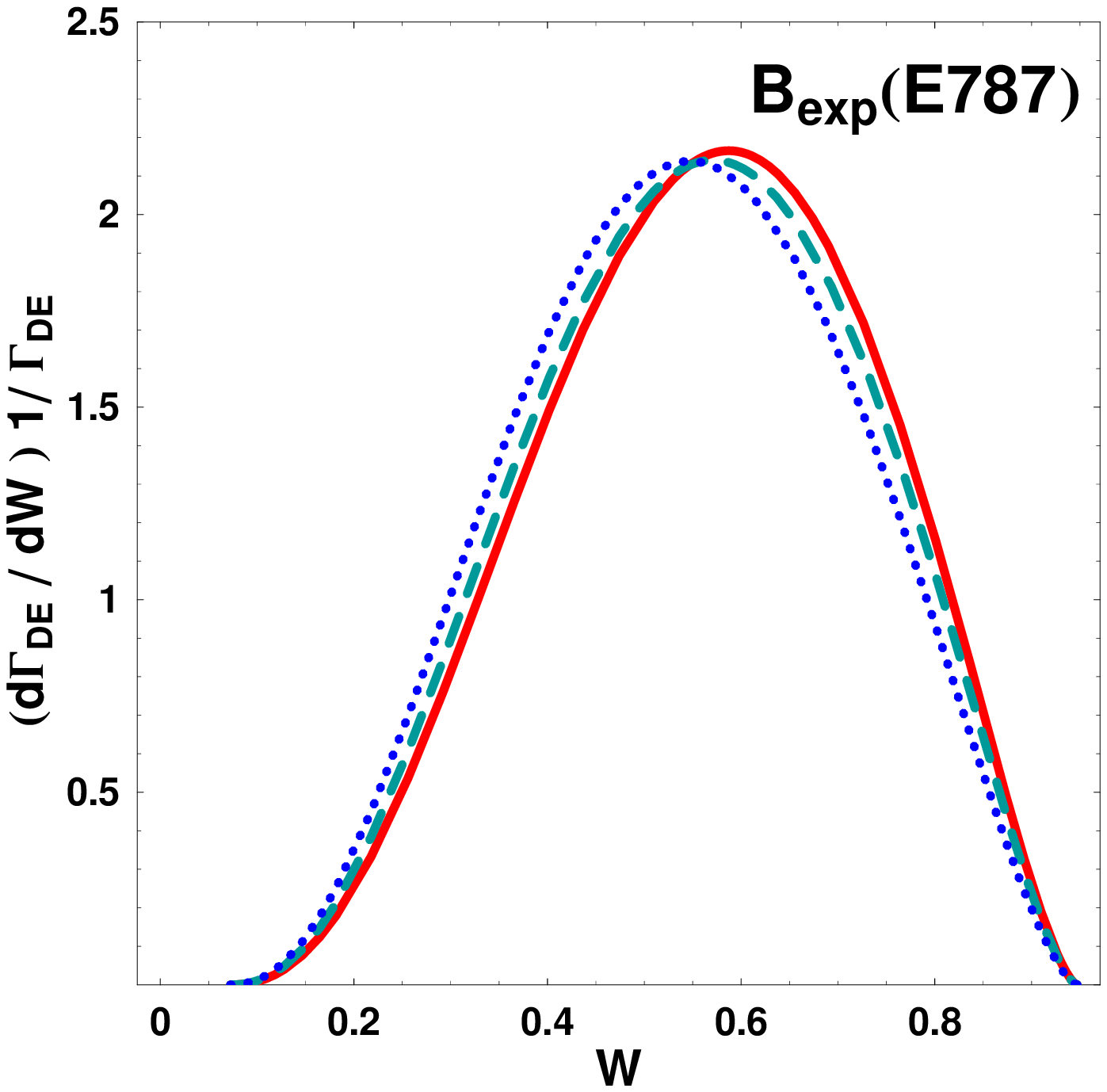}
\epsfysize=6cm\epsfxsize=7cm\epsfbox{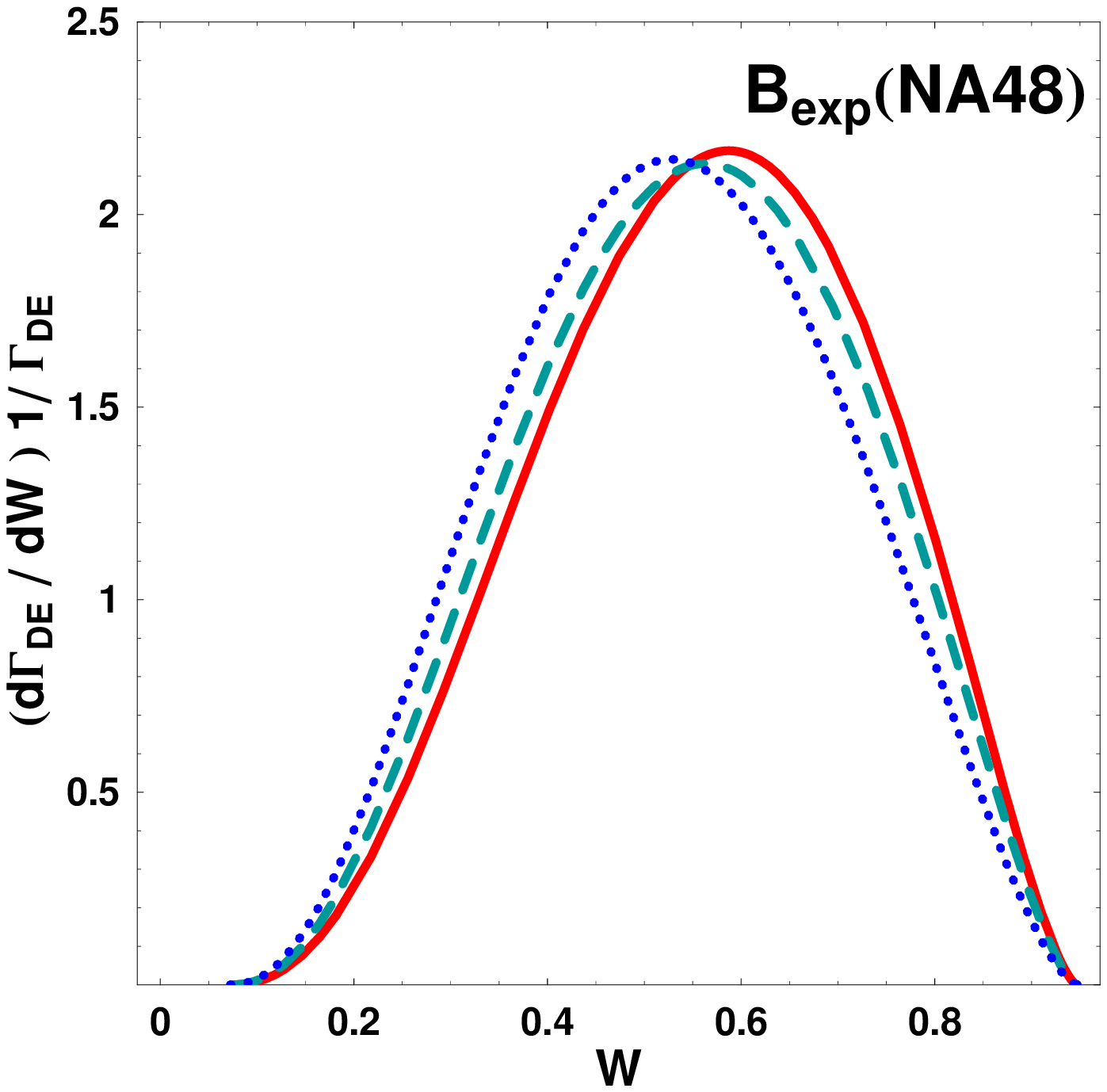}
\smallskip
\parbox{7cm}{\caption{Normalized $W$-spectra ($T_{c}^{\ast }\in \left[0, 80\right]$ MeV) for the DE magnetic
amplitude.
The upper and lower figures correspond  to the  E787 branching ratio \cite{BNL00} in TAB. \ref{tab:kppg}
 (the central solid  line of FIG. 2)  
and to the  NA48/2 branching ratio in TAB. \ref{tab:kppg} (the lower solid  line of FIG. 2) respectively.
The solid curves corresponds to a \red{constant} amplitude, while
the {\color{verdel} dashed} and {\color{blue} dotted}  curves correspond to form factors with
{\color{verdel}  $\eta_V=0.5$} and {\color{blue} $\eta_V=1.5$} respectively.
}}\label{fig:Wspect080}
\vskip.25cm
\end{figure}

As we can see the changing of the  value of the branching ratio generates a substantial effect  in   the $W-$spectra in  FIGS. 5 and 6.

Subtracting the IB contribution to the $W$-spectrum in (\ref{dWEM}) and
assuming  $E$  and $M$ constant 
one obtains
\begin{eqnarray}
 \Frac{d \Gamma (E,M)}{d W}\propto INT(E) \ W^2 +DE(E,M)\ W^4.\label{eq:fitEM}
\end{eqnarray}
Then we can fit this to the experimental data  determining  
$E$ from the interference term, $ INT(E)$, and $M$ from $DE(E,M)$. Since $M_+^{\rm pole}(z_i)$ in (\ref{KPVMD}) is obviously  not constant,
we would like to question whether the presence of this form factor  
(and  no interference), could
simulate an interference term in (\ref{eq:fitEM}) as observed by NA48/2 \cite{NA48/2,RaggiCundy}. 
In fact we plot in FIG.6 the difference 
among  normalized W-spectra :  
\begin{eqnarray}
\Delta&=&\label{eq:diffMff}\\&&
\hspace{-1cm}\Frac{1}{{\cal N}_{ff}}
\Frac{d \Gamma (0,M_+(z_i))}{d W}-
\Frac{1}{{\cal N}_{\rm const.}}
\Frac{d \Gamma (0,M_+^{\rm const.})}{d W}\nonumber
\end{eqnarray}
where the first distribution is generated by the form factor structure in eq. (\ref{MVMD}) and the second one
 by constant magnetic term. We study this difference in FIG.6 for DE  amplitudes corresponding to  form factors  with
$\eta_V=0.5$ (dashed line) and $\eta_V=1.5$ (dotted line). ${\cal N}_{ff}$ and  ${\cal N}_{\rm const.}$
are the normalization factors for   W-spectra, corresponding to the  form factor and constant magnetic term
 respectively.
\begin{figure}[hb]
\vskip.25cm\centering
\epsfysize=6cm\epsfxsize=7cm\epsfbox{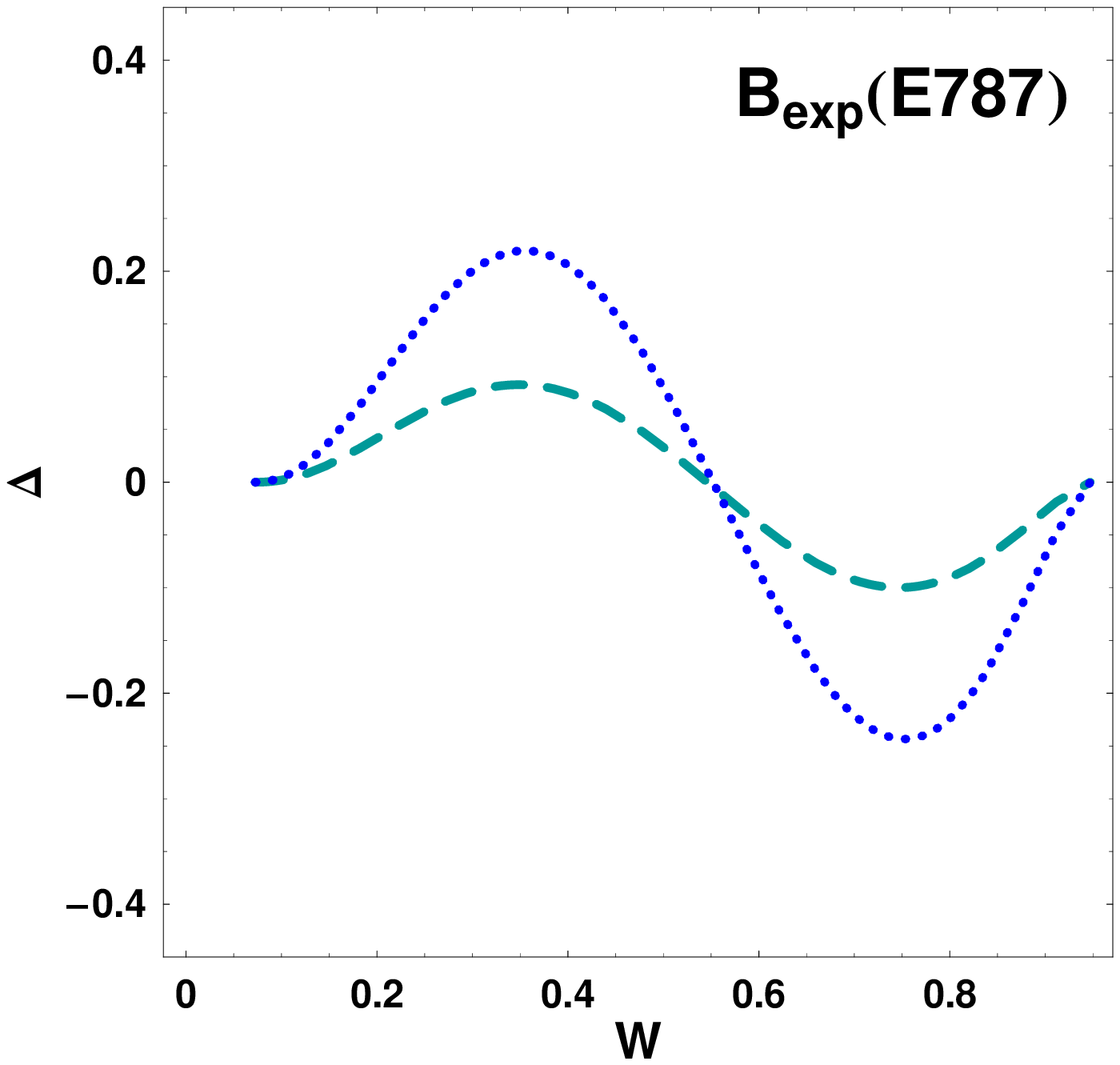}
\epsfysize=6cm\epsfxsize=7cm\epsfbox{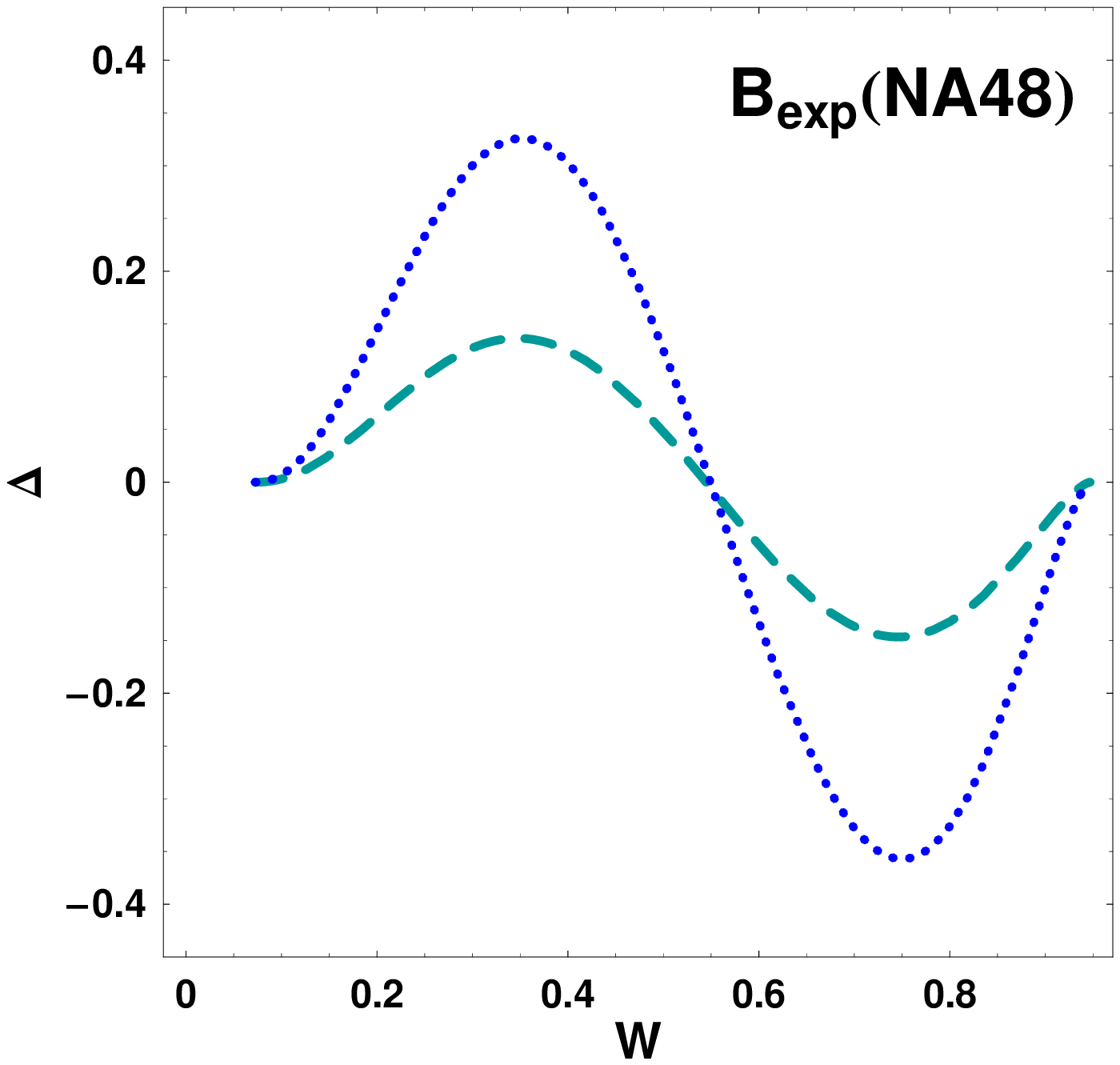}
\smallskip
\parbox{7cm}{\caption{
Plots for the quantity $\Delta$ in eq. (\ref{eq:diffMff}), obtained by subtracting the normalized 
$W$-spectrum of  constant DE amplitude from the normalized
$W$-spectra of DE  amplitude corresponding to  form factors  with
{{\color{verdel}$\eta_V=0.5$ (dashed line)}} and {{\color{blu}$\eta_V=1.5$ (dotted line)}};
upper and lower plots refer to E787 branching ratio \cite{BNL00} in 
 TAB. \ref{tab:kppg}
 (the central solid  line of FIG. 2)  
and to the  NA48/2 branching ratio in   TAB. \ref{tab:kppg}
respectively (the lower solid  line of FIG. 2).
 $T_{c}^{\ast }$-range is   $\left[0, 80\right]$ MeV.
}}\label{fig:diffWspectnew}
\vskip.25cm
\end{figure}

An interference term would appear as a line starting from the origin and going to negative values for negative interference.
It is clear that there is a correlation between the measurement of the interference and direct emissiom with form factor.
We can see  from FIG.6 that this depends also on  the size of the branching ratio: 
the effect is  $\le 20\%$ for the  E787 branching ratio  and even $ 30\%$ for 
the  NA48/2 branching ratio in   TAB. \ref{tab:kppg}  and for the extreme value of the VMD parameter, $\eta _V=1.5$ (dotted line). 
 However we can exclude that the
form factor can  completely account for the 
 interference effect.

\section{Conclusions}
VMD has played a very crucial role for our understandings of chiral dynamics and its  predictive power in the strong sector;
we expect a similar  success in the weak sector, however the path is apparentely more complicated. 
The magnetic contribution to 
$K_L\to \pi ^{+}\pi ^{-}\gamma$ decays is 
one of the few examples in the weak sector where there is phenomenological evidence of VMD.
 This is particularly interesting
for the interplay with CP violating amplitudes and  possible New Physics searches.
Furthermore, the precision level of CPT tests is such that even  $A(K_L\to \pi ^{+}\pi ^{-}\gamma)_{ DE}$ has to be  known 
with good accuracy  \cite{Ambrosino:2006ek}.

We have studied in this paper
the correlated channel $K^+\to \pi ^{+}\pi ^{0}\gamma$, in connection with the upcoming NA48/2 results.
We have analyzed in Section II a set of kinematical variables, Christ's variables, used  
also by NA48/2. 

We have numerically analyzed  how to find experimental evidence of VMD in this channel;
compared to previous literature \cite{D'Ambrosio:2000yc} we have payed particular attention to the dependence
of the parameter  $\eta_V$ and  $A^+$, entering in this study, on the measured $B(K^+\to \pi ^{+}\pi ^{0}\gamma)$ (see
 FIG. 2). After this analysis we have  concentrated on the dependence of the spectra on VMD  
 see FIGS. 3-5. 

We have explored also the possible dependence of the observed interference effect, due to electric transitions,
 on the presence of a form factor in magnetic transitions. Since  electric transitions are important from a chiral dynamics point of view this question is 
relevant: we  have found
that {\it i)} there is a correlation between  the size of the branching ratio and the shape of the spectra and {\it ii)}
 for an
accurate determination
 of the interference contribution the size of the form factor is relevant.

\vspace{1cm}
\section*{ACKNOWLEDGEMENTS}
This work has been supported in part by the European Commission (EC) RTN
FLAVIAnet under Contract No. MRTN-CT-2006-035482 and by the Italian Ministry of Education and Research (MIUR), project 2005-023102.
We thank Gino Isidori, Donald Cundy, Mauro Raggi and Marco Sozzi for continuous discussions and help.

\end{document}